\documentclass[prd,nofootinbib,superscriptaddress,twocolumn,showpacs]{revtex4}
\usepackage{amsfonts}
\usepackage{amsmath}
\usepackage{amssymb}
\usepackage{latexsym}
\usepackage{color}
\usepackage{colordvi}
\usepackage{epsfig}
\usepackage{array}
\usepackage{pifont}

\newcommand{\beq}{\begin{eqnarray}}
\newcommand{\eeq}{\end{eqnarray}}

\newcommand{\be}{\begin{equation}}
\newcommand{\ee}{\end{equation}}
\newcommand{\lwrsim}{\raise0.3ex\hbox{$<$\kern-0.75em\raise-1.1ex\hbox{$\sim$}}}
\newcommand{\etabar}{\overline{\eta}}
\newcommand{\rhom}{\overline{\rho}}

\def\Am#1#2#3{\widetilde A_{#1}^{#2}(#3)}

\def\C2#1#2{({\cal C}_2)_{#1}^{#2}}

\def\assylarge#1{\begin{array}{c} \rule[0.45cm]{0cm}{0cm} \to \\ #1 >> 1
                 \end{array}}
\def\eq#1{Eq. (\ref{#1})}

\begin{document}

\preprint{UHU-10-04}

\title{Are the low-momentum gluon correlations semiclassically determined?}

\author{Ph. Boucaud}
\affiliation{Laboratoire de Physique Th\'eorique~\footnote{Unit\'e Mixte
de Recherche du CNRS - UMR 8627}, U. Paris-XI, Bat. 210, 91405 Orsay, France.}
\author{F. De Soto}
\affiliation{Dpto. de F\'{\i}sica At\'omica, Molecular y Nuclear; 
Universidad de Sevilla, Apdo. 1065, 41080 Sevilla, Spain}%
\author{A. Le Yaouanc}
\affiliation{Laboratoire de Physique Th\'eorique~\footnote{Unit\'e Mixte
de Recherche du CNRS - UMR 8627}, U. Paris-XI, Bat. 210, 91405 Orsay, France.}
\author{J. Rodr\'{\i}guez--Quintero}
\affiliation{Dpto. de F\'{\i}sica Aplicada; 
Fac. Ciencias Experimentales, Universidad de Huelva, 21071 Huelva, Spain}



\begin{abstract}
\bigskip
We argue that low-energy gluodynamics can be explained in terms of semi-classical 
Yang-Mills solutions by demonstrating that lattice gluon correlation functions fit to 
instanton liquid predictions for low energies and, after cooling, in the whole range. 
\end{abstract}

\maketitle

\section{Introduction}

Semi-classical methods exploiting the non-dispersive solutions of non-linear classical field equations, 
named solitons, are widely used in physics. In particular, the time-dependent four-dimension solutions 
of the Yang-Mills equations (instantons, merons, monopoles, etc.) 
have been proposed to explain the low energy properties 
of quantum chromodynamics (QCD) respecting, for instance,
the lower part of the Dirac Operator Spectrum or 
the confinement problem~\cite{Schafer:1996wv,diakonov,verb,hutter,Glimm:1977sx}.
The interest on a semi-classical understanding of confinement 
has been recently renewed by lattice studies~\cite{Lenz:2003jp}.

Other recent lattice results~\cite{Boucaud:2002fx} point to the 
semiclassical nature of gluon correlations in the low energy regime of 
QCD. Instanton properties are traditionally measured on the lattice
based on a geometrical localisation of instanton shapes after 
a cooling procedure that eliminates quantum fluctuations. 
This method incorporates a number of known biases,
such as instanton disparition and distortion, etc. 
Alternatives used by several authors are to
fix a given number of cooling sweeps (where
authors assume distortions are not 
significant)~\cite{smith} or to study
the evolution of properties with cooling
(allowing to extrapolate properties to
the original --uncooled-- situation)~\cite{Negele:1998ev}.

During last years, topological properties of QCD 
have been measured directly from Dirac operator
spectrum, made possible with the use of
improved Gispard-Wilson fermions~\cite{Horvath:2003yj}.
On the other hand, other method to look for
instantonic properties that does not require of
a cooling procedure arises from the study 
of gluon correlation functions in 
the quenched approximation,
whose non perturbative part can be nicely described in terms 
of instantons~\cite{Boucaud:2002fx}.
In this note, we will present some aditional results
concerning this method, that confirm the latter hypothesis
that gluon correlations can be described semiclassically.

\section{Instanton Liquid picture}

Recalling some formulae from previous papers, our
aim is to study the gluon correlations, that can be 
traced through the following euclidean scalar form factors:

\beq
G^{(3)}(k^2) = \frac{-i}{18 k^2} \frac{f^{a_1 a_2 a_3}}{24} 
\langle \Am{\mu_1}{a_1}{k_1} \Am{\mu_2}{a_2}{k_2} \Am{\mu_3}{a_3}{k_3} 
\rangle \nonumber \\
\times \left( T^{tree}_{\mu_1 \mu_2 \mu_3} + 
\frac{(k_1-k_2)_{\mu_3} (k_2-k_3)_{\mu_1} (k_3-k_1)_{\mu_2}}{2 k^2} \right) \\
G^{(2)}(k^2) =  \frac{\delta^{a b}}{24} 
\left(\delta_{\mu \nu}-\frac{k_{\mu} k_{\nu}}{k^2} \right) \ 
\langle \Am{\mu}{a}{k} \Am{\nu}{b}{-k} \rangle \ .
\label{G2y3}
\eeq
These are the non-perturbative MOM definitions of two- and three-gluon Green function in Landau
gauge,  where $A_\mu^a$ is the gluon gauge field,  $T^{tree}$ is the standard tree-level tensor
for the three-gluon vertex, $f^{a_1 a_2 a_3}$ stands for the SU(3) structure  constants and the
renormalization point is defined by $k_1^2=k_2^2=k_3^2=k^2$ and $k_1+k_2+k_3=0$.

On the other hand, the gauge field in the instanton picture 
and within the sum-ansatz approach, can be written as
\beq
g A_\mu^a= 2 \sum_i R^{a \alpha}_{(i)} \etabar_{\mu\nu}^\alpha \frac{ \left(x^\nu-z_\nu^i \right)}{|x-z^i|^2} 
\phi\left(\frac{|x-z^i|}{\rho_i} \right) \  ,
\label{Bsu2}
\eeq
where $g=(6/\beta)^{1/2}$ is the bare gauge coupling in terms of the lattice parameter $\beta$,
$\etabar$ is known as 't Hooft  symbol and $R^{a \alpha}$ represents the color rotations
embedding the canonical SU(2) instanton solution in the SU(3) gauge  group, $\alpha=1,\cdots,3$
($a=1,\cdots,8$) being an SU(2) (SU(3)) color index. The sum is extended over all the instantons
and  anti-instantons (we should then replace the 't Hooft symbol $\etabar$ by $\eta$) in the
classical background of the  gauge configuration. $\phi(x)$ is the instanton profile function.

If we neglect instanton position and color correlations, eqs. (\ref{G2y3},\ref{Bsu2}) lead to 
\beq
G^{(m)}(k^2) =  n \frac{4 k^2}{m} \left( \frac{\beta}{96 k^2} \right)^{m/2}  <\rho^{3m}  I(k \rho)^m > \ ,
\label{green}
\eeq
for $m=2,3$; where $n$ is the instanton density. It depends on
the functional $I(k\rho)$ of the general instanton profile, $\phi(x)$, 
\beq
I(s) \ = \frac{8 \pi^2}{s} \int_0^\infty\ z dz  J_2(sz) \, \phi(z) \ ,
\eeq
and $< \ \cdots \ >$ means the average over instanton sizes with a given normalised instanton radius 
distribution, $\mu(\rho)$.

Of course, Green functions in the instanton liquid picture depend on the
instanton profile and  radius distribution and both are indeed far from being
definitively known. As far as one searchs for  some striking traces of a
semi-classical picture success describing gluon correlations, the  necessity
of those two inputs disturbs our purposes. In \cite{Boucaud:2002fx}, we
solved this  inconvenience not by directly studying two- and three-gluon
Green functions but a proper ratio of them,
\beq\label{alpha}
\alpha_{\rm MOM}(k) \ = \ \frac{k^6}{4 \pi} \frac{\left(G^{(3)}\right)^2}{\left(G^{(2)}\right)^3} 
= \ \frac{k^4}{18 \pi n} \ \frac{< \rho^9  I(k\rho)^3 >^2}
{< \rho^6  I(k\rho)^2 >^3} \ ,
\eeq
defining the non-perturbative QCD coupling constant 
in the symmetric MOM renormalization scheme. Only by 
assuming a narrow instanton radius distribution, the 
instanton liquid model approximatively predicts a $k^4$-power 
law for the coupling in \eq{alpha}. 
Lattice estimates of the QCD coupling constant clearly manifest 
such a $k^4$-power behaviour for the low-momentum 
range~\cite{Boucaud:2002fx}, although this pattern is distroyed by 
quantum interferences above $k\simeq 1$ GeV. 
The latter nicely suggests that 
the larges-distances gluon correlations are dominated by the 
classical solutions of the QCD lagrangian. However, a rather firm 
confirmation of this conclusion is to be obtained after eliminating 
quantum fluctuations from the lattice gauge configurations 
(by a cooling procedure) and recovering all over the range the classical 
pattern.

Moreover, the asymptotic behavior
\beq
I(s) \assylarge{s} \frac {16 \pi^2}{s^3} \ ,
\eeq
is obtained with the only condition~\footnote{Satisfied just by requiring 
a well-defined Yang-Mills  action, that is, independent of instanton profile
and even true for merons.} of $\phi(0)=1$. Thus,
\beq\label{Gmfin}
< \rho^{3m} I(k\rho)^m > & \xrightarrow[k\rho\gg 1]{}& \left(\frac{16 \pi^2}{k^3}\right)^m \ \ 
\mbox{\rm and} \\
G^{(m)}(k^2) & \xrightarrow[k\rho\gg 1]{} &  n \frac 4 m \left(\frac{8 \beta}{3} \right)^{m/2} k^{2-4m} \ .
\nonumber 
\eeq
Therefore, the large-momentum limit do not depend on the radius 
distribution nor the instanton profile (the inclusion of 
merons in the classical background would neither modify the results in 
\eq{Gmfin}) and can provide a much cleaner indication of instanton 
effects. The question about this limit will be adressed in the 
following.

\section{Cooling}

After Montecarlo integration, lattice gluon field
configurations are dominated by short-distance 
correlations. No trace of instantons can
be (in principle) seen except, as suggested by obtaining the 
$k^4$-power law in \cite{Boucaud:2002fx}, for the deep-infrared 
domain where large-distance correlations should be dominant. 
A method widely used
to detect instantons in lattice  configurations
is to perform a ``cooling'' 
procedure~\cite{Teper:1985rb,Schafer:1996wv} that eliminates
higher energy modes and, as has been extensively
shown, reveals instanton-like structures.

The procedure, proposed by Teper~\cite{Teper:1985rb}, 
lies on an iterative change
of the lattice variables by the values that
locally minimises the action. This procedure
progressively reduces the action, suppressing
quantum correlations in the field configurations.

Cooling methods have been criticised as they modify 
the configuration (instanton sizes become distorted, instanton and anti-instanton 
anhilate to each other, ...) and no indication is given about
how to recover the physical information. Some gain on this problem
was given by Ringwald {\it et al.}~\cite{Ringwald:1999ze}, that proposes
that cooling acts differently according to the lattice spacing
used. They claim that the effect of cooling grows as the correlations
it introduces, with the square root of the number of sweeps. Moreover
it allows to indicate the equivalent number of cooling sweeps for
different lattice spacings in terms of the so called
cooling radius:
\beq\label{rc}
r_c=a\sqrt{n_c} \ ,
\eeq
where $a$ is the lattice spacing and $n_c$ the number of cooling
sweeps. Thus, any property of the instanton configuration can be 
traced through the cooling procedure as a function of the cooling 
radius and extrapolated back to the undistorted 
configuration.

In the next section we will use Teper procedure to suppress the short-distance 
correlations and force the underlying picture from large-distance correlations to emerge. 
We also test the evolution of the different lattice configurations 
with cooling in terms of the cooling radius.

\section{Lattice Results}

\begin{figure}[h!]
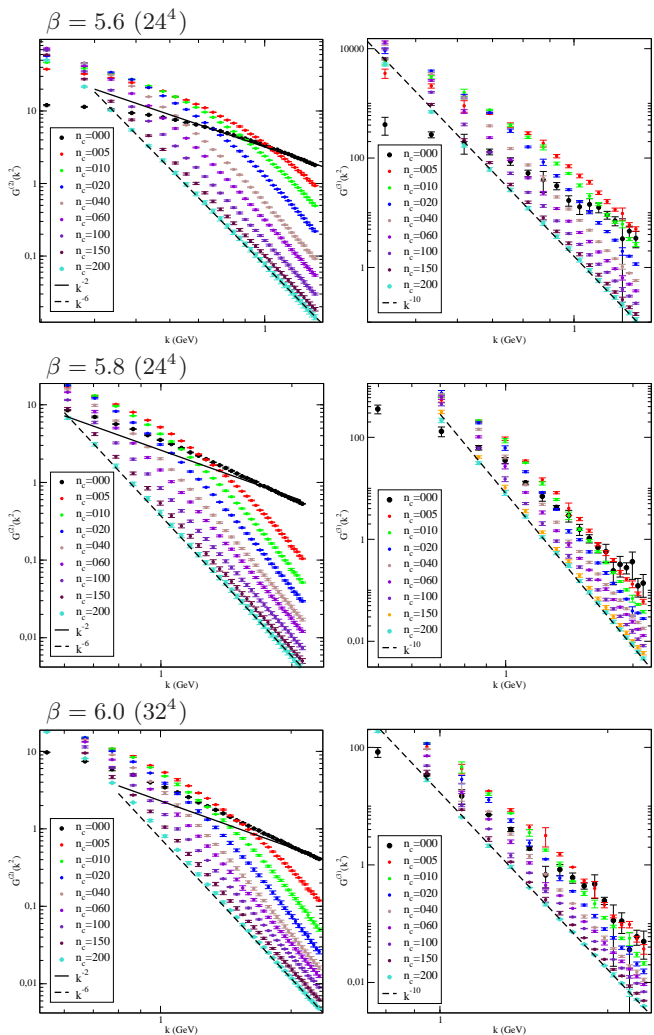

\begin{center}
\begin{tabular}{cc}
$\beta=5.6 \ (24^4)$ \rule[0cm]{1.2cm}{0cm} & \\
\includegraphics[width=10pc]{propag2456.eps} &
\includegraphics[width=10pc]{vertex2456.eps} 
\\
$\beta=5.8 \ (24^4)$ \rule[0cm]{1.2cm}{0cm} & \\
\includegraphics[width=10pc]{propag2458.eps} &
\includegraphics[width=10pc]{vertex2458.eps} 
\\
$\beta=6.0 \ (32^4)$ \rule[0cm]{1.2cm}{0cm} & \\
\includegraphics[width=10pc]{propag3260.eps} &
\includegraphics[width=10pc]{vertex3260.eps} 
\end{tabular}
\caption{\small{\it Evolution of $G^{(2)}(k^2)$ (left) and $G^{(3)}(k^2)$ (right) 
with cooling, a clear transition happens after a few cooling sweeps.}}
\label{Fig1}
\end{center}
\end{figure}

We have performed simulations for $\beta=5.6,5.8$ on a $24^4$ lattice and 
for $\beta=6.0$ on a $32^4$, obtained the gauge field Montecarlo configurations 
and computed the two- and three-gluon Green functions. Then, we applied the 
Teper procedure discussed in the previous section to {\it ``cool''} the 
field configurations and again computed the Green functions for different numbers 
of cooling sweeps. We present the results in the plots of fig. \ref{Fig1} and, 
in the following, we will discuss them.

\subsection{Green functions}

The first and more impressive hint we observe in the plots of fig. \ref{Fig1} 
is that both gluon propagator and vertex join the asymptotic classical behaviour predicted 
by \eq{Gmfin} after a few cooling sweeps. In particular, for the gluon propagator 
(left plots), while in perturbation theory it decreases at high momentum, $k$, with 
$k^{-2}$ (except for the anomalous dimension logarithm), after killing quantum correlations 
we obtain a rather clean $k^{-6}$ behaviour.
For the three-gluon vertex (right plots), correspondingly, we observe $k^{-10}$
instead of $k^{-6}$ after cooling. These observations allow 
to conclude that after cooling, correlations at high energies
are dominated by instanton-like structures.

A second remarkable hint is how the asymptotic classical regime is reached: the higher is 
the number of cooling sweeps, the lower are the energies where both Green functions 
begin to behave as given by \eq{Gmfin}. As a matter of fact, the typical instanton size, 
$\rho$, grows with the cooling and the asymptotic formula \eq{Gmfin} stands for 
$k \gg \rho^{-1}$. 
Furthermore, if we compare lattice data at different $\beta$ 
for the same number of cooling sweeps, we also qualitatively observe that the effect of cooling grows 
with the lattice spacing, $a(\beta)$, as expected after considering the cooling radius in \eq{rc} 
(the lower is $\beta$, the lower are the energies where the asymptotic behaviour is reached).

\subsection{Evolution with cooling}

The effect of the cooling in terms of the cooling radius can be quantitatively analyzed 
because the instanton density can be extracted from a fit of the lattice
data in the asymptotic regime to \eq{green} for all the lattices and number of cooling sweeps. 
The asymptotic behaviour does not depend on the details of the
instanton liquid, and is the same for any instanton profile
or even for merons \cite{Lenz:2003jp}. Thus, the fitted density should not be sensitive to the 
possible instanton deformations induced by the cooling.

\begin{figure}[!h]
\begin{center}
\includegraphics[width=18pc]{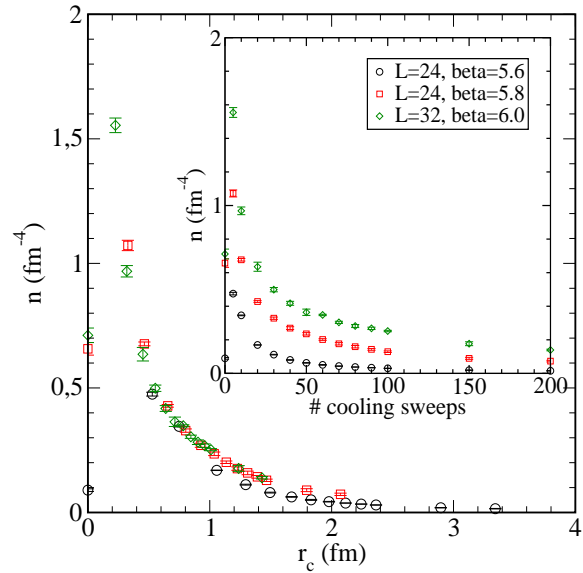}
\caption{\small \it Density obtained from a fit of the lattice data for $G^{(3)}(k^2)$
to the classical expresions (\ref{Gmfin}).  Different curves for different values of
$\beta$ (detail) do coincide in terms of cooling radius.}
\label{density}
\end{center}
\end{figure}

In fig. \ref{density}, we present the evolution of the density with the number of cooling sweeps and 
with the cooling radius. The nice matching of the results for the three lattices proves that the 
estimated density through the fit to \eq{Gmfin} indeed evolves with the correlation length introduced 
by the cooling. In other words, for the same cooling correlation length all our {\it cooled} lattice 
data match to each other and behave as classically predicted in \eq{Gmfin}.

\subsection{The running coupling}

A final hint of plots in fig. \ref{Fig1} to be discussed is why the cooling procedure, supposed to 
destroy only short-distance correlations, immediately modify the low-momentum range of Green 
functions. This is because of the density damping induced by the cooling. The instanton profile 
depends of the instanton density~\cite{Diakonov:1983hh,Boucaud:2002fx} and the low-momentum Green 
functions are affected by instanton profile effects~\cite{Boucaud:2002fx}, diverging as 
$k \to 0$ in the zero-density limit. 

A simple manner of testing this explanation, with no 
detailed discussion about the instanton profile, is analyzing the QCD running coupling in \eq{alpha}.
As we above discussed, if the radius-distribution width is small enough, the instanton-profile effects
from two- and three-gluon Green function very approximatively cancel to each other 
in \eq{alpha} and we recover a $k^4$-power law. We can write:

\beq\label{alpha2}
\alpha\left(k^2\right) \ = \ \frac 1 {18 \pi n \rhom^4} \times \left\{
\begin{array}{ll}
c \left( k \rhom\right)^{4-\varepsilon} & k\rhom \ \lwrsim \ 1 \\
(k \rhom)^4 & k\rhom \gg 1
\end{array} \right. \ .
\eeq
The low-momentum limit is taken from~\cite{Boucaud:2002fx}, where $c\simeq 1.6$ and 
$\varepsilon \simeq 0.1-0.2$. Both were estimated for $0.4 < k < 0.9$ GeV with a 
instanton radius distribution taken from \cite{Diakonov:1983hh} centered at $\rhom=1.5$ GeV 
and are rather independent 
of the choice for the particular profile function.

\begin{figure}[h!]
\begin{center}
\includegraphics[width=16pc]{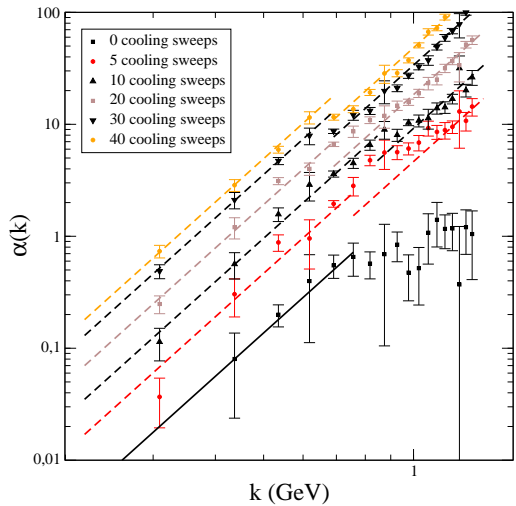}
\caption{\small \it Log-log plot of the running coupling defined in \eq{alpha} in terms of 
momenta. The two limits predicted by \eq{alpha2} are shown by straight and dashed lines
respectively.}
\label{Fig3}
\end{center}
\end{figure}

In fig. \ref{Fig3}, the observance of the $k^4$-power law by the running coupling defined in
\eq{alpha} for both low and large-momentum range is manifest. We present, as an example, 
the results from the simulation at $\beta=5.6$ in a $24^4$ lattice, but the same is obtained 
from the other simulations. For the three lattices, the shifts of the straight line fitting 
the low-momentum data respect to the one fitting in large momenta correspond to 
$c \simeq 1.5-2.0$ defined in \eq{alpha2}, in good agreement with results in~\cite{Boucaud:2002fx}.
Consequently, densities plotted in fig. \ref{density}, obtained by fitting lattice data to the 
asymptotic formula \eq{Gmfin} for large momenta, correspond in practice to the same fitted for 
low momenta in~\cite{Boucaud:2002fx}.

\section{Conclusions}

In conclusion, after suppression of the short-distance correlations by cooling, the dominance of 
the semi-classical large-distance correlations leads to a nice description of the asymptotic 
behaviour of gluon Green functions within the instanton picture. In the low momentum range 
($k \ \lwrsim \ 1$ GeV), the cooling procedure only modifies details of the Instanton liquid like 
density or instanton profile. This can be seen by analyzing the QCD running coupling defined 
by \eq{alpha} which shows the same classical pattern for both low and large momenta, as predicted 
by \eq{alpha2}. Consequently, the low-momentum gluon Green functions from Montecarlo gauge 
configurations are determined by the classical correlations, although their description within the 
Instanton picture requires to know the details of the Instanton liquid.

\medskip

We are especially indebted to O.~P\`ene, J.P.~Leroy and J.~Micheli for comments and 
suggestions very valuable for the ellaboration of this letter. This work has been partially 
supported by a grant of the spanish Ministery of Education and Science.

\end{document}